\documentstyle[epsfig,aps,prb,multicol]{revtex}
\def\beq{\begin{equation}}
\def\eeq{\end{equation}}
\def\beqa{\begin{eqnarray}}
\def\eeqa{\end{eqnarray}}
\def\p{\varphi}
\tightenlines
\begin{document}
\title{On the derivation of a high-velocity tail from the 
Boltzmann--Fokker--Planck equation for shear flow}
\author{L. Acedo,\footnote{Departamento de F\'{\i}sica,
Universidad  de  Extremadura, 
E-06071 Badajoz, Spain
}
A. Santos,$^\dag$ and A. V. Bobylev\footnote{Division of Engineering 
Sciences, Physics and Mathematics,
Karlstad University, Karlstad, Sweden}}
\date{\today}
\maketitle

\begin{abstract}
Uniform shear flow is a paradigmatic example of a
nonequilibrium { fluid} state exhibiting non-Newtonian behavior. 
It is characterized by uniform density and temperature
and a linear velocity profile  $U_x(y)=a y$, where $a$ is the constant  shear rate. 
In the case of a rarefied gas, all the relevant physical information is
represented by the one-particle
 velocity
distribution function $f({\bf r},{\bf v})=f({\bf V})$, with ${\bf V}\equiv {\bf
v}-{\bf U}({\bf r})$, which satisfies the standard nonlinear
integro-differential Boltzmann equation.
We have studied this state for a
two-dimensional gas of Maxwell molecules with a
 collision rate $K(\theta)\propto\lim_{\epsilon \to 0}
\epsilon^{-2}\delta(\theta-\epsilon)$, where $\theta$ is the scattering angle,
in which case the nonlinear Boltzmann collision operator reduces to a
Fokker--Planck operator. We have found analytically that for shear rates
larger than a certain threshold value $a_{\text{th}}\simeq 0.3520\nu$ (where $\nu$ is an
average collision frequency and $a_{\text{th}}/\nu$ is the real root of the cubic equation
$64x^3+16x^2+12x-9=0$) the velocity distribution function 
exhibits { an algebraic} high-velocity tail of the form $f({\bf V};a)\sim 
|{\bf V}|^{-4-\sigma(a)}\Phi(\p; a)$, where $\p\equiv \tan V_y/V_x$ and  
the angular
distribution function $\Phi(\p; a)$ is the solution of a
modified Mathieu equation. The enforcement of the periodicity condition 
$\Phi(\p; a)=\Phi(\p+\pi; a)$
allows one to obtain the
exponent $\sigma(a)$ as a function of the shear rate. It 
diverges when $a \rightarrow a_{\text{th}}$ and tends
to a minimum value $\sigma_{\text{min}} \simeq 1.252$ in the limit $a\to\infty$.
As a consequence of this power-law decay for $a>a_{\text{th}}$, all the velocity moments of
a degree equal to or
larger than $2+\sigma(a)$ are divergent.
In the high-velocity domain the velocity distribution is highly anisotropic,
with the angular distribution sharply concentrated around a preferred
orientation angle $\widetilde{\p}(a)$, which rotates from
$\widetilde{\p}=-\pi/4,3\pi/4$ when $a\to a_{\text{th}}$ to $\widetilde{\p}=0,\pi$ in
the limit $a\to\infty$. 
\end{abstract}

\draft


\section{Introduction}

\label{sec1}
In the investigation of the physical properties of fluids far from
equilibrium, one usually { focuses} on the nonlinear dependence of the 
momentum and
heat fluxes on the gradients of the hydrodynamic fields.
The associated transport properties are related to the population of molecules with
energies of the order of or less than the mean kinetic energy, so that molecules
moving with velocities much larger than the thermal velocity hardly contribute
to those properties. However, the knowledge of the high-energy population in
nonequilibrium states is important not only from a theoretical point of view
but also because that population may play a crucial role in processes such as
chemical reactions with a high activation energy or in the controlled
thermonuclear fusion of a confined hydrogen plasma.

Of course, a general description of the high-energy population for arbitrary
nonequilibrium states is not possible. { Therefore, it is 
worthwhile gaining some insight} 
 by considering particular states.
It can be fairly said that one of the most extensively
studied nonequilibrium states  is the so-called uniform shear flow (USF).
At a macroscopic level, it is characterized by a constant density
$n$, a uniform temperature $T$, and a linear profile of the $x$ component 
of the flow velocity along the $y$ direction, i.e., $U_x(y)=a y$, $a$
being the constant shear rate. This shear rate represents the only
control parameter needed to measure the departure of the USF state from
equilibrium.
At a microscopic level, \cite{DBS86} the USF is described by a 
solution of the Liouville equation with Lees-Edwards boundary 
conditions, \cite{LE72} which can be seen as periodic boundary 
conditions in the {\em local} rest frame.
These conditions assure the consistency of uniform shear, density,
and temperature, even far from equilibrium.
This state has been widely used to study rheological properties, such as
shear thinning and viscometric effects. 
It must be { borne} in mind that, except in the linear regime,
the USF is not equivalent to  the planar Couette flow. 
In the latter, the shearing is produced by ``realistic'' walls in relative
motion, so that the 
boundary conditions correspond to particles interacting with the walls rather
than to the generalized periodic Lees-Edwards boundary conditions.
In contrast to USF,  boundary effects are present in the Couette flow and, in
addition, the shear rate, 
density, and temperature are local quantities.\cite{MGS01}
Far from equilibrium, the rheological properties of the Couette flow
differ from those of the USF. \cite{SGB92}

In general, no rigorous theory based on first principles exists for the
USF state. On the other hand, if one restricts oneself to the case of dilute gases,
the most relevant physical information is contained in the one-particle
velocity distribution function $f({\bf r},{\bf v},t)$ and then the Liouville
equation or, equivalently, the BBGKY hierarchy can be successfully 
contracted to
a closed equation for $f$, namely
 the nonlinear Boltzmann equation. \cite{C75,C90} A number of exact results 
 have
 been derived from  the Boltzmann equation specialized to the case of Maxwell
 molecules under USF.
Almost half a century ago, Ikenberry and Truesdell\cite{IT56,T56,TM80} showed that the infinite 
hierarchy of moment equations could be recursively solved.
In particular, they obtained the temporal evolution of the second-degree 
moments, 
which are the quantities related to the rheological properties, for arbitrary
values  of the shear rate.
Truesdell and Muncaster \cite{TM80}  analyzed the temporal evolution of the
third-degree moments (for instance, the heat flux), which are expected to vanish for long times because of
symmetry. They observed that for shear rates larger than a certain value some
of those moments increased with time, a feature they referred to as an {\em
instability\/} in the heat flux solution. However, further analysis \cite{SG95}
has proved that such an increase is actually  a consequence of the viscous
heating and that the heat flux does indeed vanish for long times when it is
properly scaled with respect to the thermal velocity, { so} the 
apparent instability { is then}
removed.
More recently,  explicit expressions for the fourth-degree moments have also 
been derived.\cite{SG95,SGBD93}
While the (scaled) second-degree moments remain finite for arbitrary shear rates, there 
exists a critical value $a_c$ of the shear rate, beyond which the 
fourth-degree
moments diverge.
The analysis of this singular behavior has been extended to moments  up to  degree 
36  and 240 for three-dimensional\cite{MSG96} and two-dimensional\cite{MSG97}
systems,
respectively. These exact results  show that  the moments of an even degree $k\geq 4$ 
are divergent if the shear rate is larger than a certain $k$-dependent critical value
that decreases as the degree $k$ increases. 
This behavior of the moments { indicates} that the distribution function
presents an algebraic high-velocity tail. { This} expectation has been
strongly supported  by direct
Monte Carlo simulations, \cite{MSG97,MGS97} { which also show that the 
high-velocity distribution function presents a strong anisotropy.} To the 
best of our knowledge { these high-velocity properties  have} not been so 
far confirmed at a
theoretical level. The aim of this paper is to
fill this gap.

Since, due to the complexity of the collision term, the nonlinear Boltzmann
equation is extremely difficult to solve,  we consider here a simplified model of two-dimensional
Maxwell molecules.
In this model the collision rate is assumed to vanish for all the scattering
angles except for a small angle corresponding to  grazing collisions. This
allows us to replace the nonlinear collision operator by a Fokker--Planck
operator\cite{C75} that, on the other hand, preserves most of the general 
features of the
original operator.
By assuming a high-velocity tail of the form $f({\bf V})\sim
V^{-4-\sigma}\Phi(\varphi)$, where ${\bf V}={\bf v}-{\bf U}$ is the peculiar
velocity and $\varphi=\tan^{-1}V_y/V_x$ is the polar angle in velocity space,
the Boltzmann equation yields a linear second-order ordinary differential
equation for $\Phi$. The periodicity condition on $\Phi(\varphi)$ determines
$\sigma$ as a function of the shear rate, thus confirming the assumed
algebraic tail. The results show that the exponent $\sigma(a)$ is a monotonically decreasing
function that diverges when $a$ approaches a threshold value $a_{\text{th}}$
and tends to a minimum value $\sigma_{\text{min}}\simeq 1.252$ in the limit of
large shear rates.
The first property means that for shear rates smaller than $a_{\text{th}}$ the
distribution function decays for large velocities more rapidly than any power
law { (e.g., as a stretched exponential)}, while the second property 
implies that, even for large shear rates, all
the moments of the form $\langle V^k\rangle$ with $k\leq
2+\sigma_{\text{min}}$ are finite. The orientation distribution $\Phi(\varphi)$
in the high-velocity domain is concentrated around a preferred angle
$\widetilde{\p}(a)$ that rotates counter-clockwise as the shear rate
increases. Moreover, this distribution is infinitely sharp for the extreme
values of the shear rate: $\lim_{a\to
a_{\text{th}}}\Phi(\varphi)=\frac{1}{2}[\delta(\varphi-3\pi/4)+\delta(\varphi-7\pi/4)]$, 
$\lim_{a\to\infty
}\Phi(\varphi)=\frac{1}{2}[\delta(\varphi)+\delta(\varphi-\pi)]$. 
To our knowledge, this is the first time that such a detailed picture of the
high-velocity behavior of the solution to the Boltzmann equation in a far from
equilibrium state has been analytically described.

The organization of this paper is as follows. The uniform shear flow state is
described in Sec.\ \ref{sec2}. Some exact scaling properties in the case of
Maxwell molecules are used to map the time-dependent state of the system onto an
equivalent ``thermostatted'' state which reaches a { 
nonequilibrium} steady state for long
times.
The simple scattering model representing grazing collisions is introduced in
Sec. \ref{sec3}. The temporal evolution of the (scaled) fourth-degree moments is
analyzed, the results showing that they diverge for shear rates equal to or
larger than a certain critical value $a_c$, as expected from previous analyses
for more realistic scattering laws.\cite{MSG97,MGS97} The derivation of the
high-velocity tail of the form $f({\bf V})\sim V^{-4-\sigma}\Phi(\varphi)$ is
worked out in Sec.\ \ref{sec4}, where the dependence of the exponent $\sigma$ on
the shear rate is obtained from the periodicity condition
$\Phi(\varphi)=\Phi(\varphi+\pi)$. Finally, the results are discussed in Sec.\
\ref{sec5}.

\section{Uniform shear flow}
\label{sec2}

The most relevant quantity to determine the nonequilibrium properties of a 
dilute gas is the one-particle velocity distribution function
$f({\bf r},{\bf v},t)$. Its time evolution  is governed by the nonlinear 
Boltzmann equation. 
In the absence of external forces, it reads: \cite{C75,C90,DvB77}

\begin{eqnarray}
\label{b1}
\frac{\partial}{\partial t}f+{\bf v}\cdot \frac{\partial}{\partial {\bf r}}f
&=&\int d{\bf v}_1 \int d\Omega \,|{\bf v}-{\bf v}_1|I(|{\bf v}-{\bf v}_1|,
\theta)(f'f_1'-ff_1) \nonumber\\
&\equiv&Q[f,f],
\end{eqnarray}
where $I(g,\theta)$ is the differential cross section ($\theta$ being the 
scattering angle) and we are using standard 
notation to denote the distribution function evaluated at pre- and 
post-collisional velocities. Of course, Eq.\ (\ref{b1})
must be supplemented with the appropriate initial and boundary
conditions.

Let us now introduce the velocity field

\begin{equation}
\label{b2}
U_i({\bf r})=a_{ij}r_j,\quad a_{ij}=a \delta_{ix}\delta_{jy},
\end{equation}
where $a$ is a constant shear rate.
We define the uniform shear flow (USF) state as the one that is spatially
homogeneous when the velocities of particles are referred to a Lagrangian
frame moving with the velocity field ${\bf U}({\bf r})$, i.e.,
\begin{equation}
\label{b3}
f({\bf r},{\bf v},t)=f({\bf V},t),
\end{equation}
where ${\bf V}\equiv {\bf v}-{\bf U}({\bf r})$ is the peculiar velocity.
Consequently, the Boltzmann equation
(\ref{b1}) becomes
\begin{equation}
\label{b4}
\frac{\partial}{\partial t}f-\frac{\partial}{\partial V_i}a_{ij}V_j f
=Q[f,f].
\end{equation}
The usual boundary conditions used to generate the USF are the Lees-Edwards
periodic boundary conditions, \cite{LE72,DSBR86} but the so-called 
``bounce-back'' boundary conditions \cite{C89,CC94} are also consistent with the 
USF.
It is worthwhile noting that Eq.\ (\ref{b4}) can be interpreted as representing 
a {\em homogeneous\/} 
state under the action of the nonconservative external force $F_i=-ma_{ij}V_j$.
Note also that Eq.\ (\ref{b4}) is invariant under the transformation 
$(V_x,V_y)\to(-V_x,-V_y)$.

Let us now  particularize to the case of Maxwell molecules, i.e., particles 
interacting via a
potential $\phi(r)\propto r^{-2(d-1)}$, where $d$ is the dimensionality of the 
system. In that case, the collision rate $gI(g,\theta)=K(\theta)$ is
independent of the relative velocity $g$. \cite{E81,B88}
{}From a mathematical point of view, this makes the Boltzmann collision operator 
more tractable than for other interaction models.
In particular, any collisional
moment of degree $k$ can be expressed as  a bilinear combination of moments of 
$f$ of degrees $k'$
and $k''$, such that $k'+k''=k$. \cite{TM80,B88,M89}
This allows one, in principle, to solve recursively the hierarchy of moment 
equations arising from Eq.\ (\ref{b4}), even though the explicit form for $f$ is 
not known. \cite{IT56,T56,TM80,SG95,SGBD93,MSG96,MSG97,MGS97,MSG96b}
In addition, Eq.\ (\ref{b4}) for Maxwell molecules exhibits the following
scaling property. \cite{SG95,DSBR86}
Let us introduce the scaled quantities
\begin{equation}
\label{b5}
\overline{\bf V}=e^{-\alpha t} {\bf V},
\end{equation}
\begin{equation}
\label{b6}
\overline{f}(\overline{\bf V},t)=e^{d\alpha t} f({\bf V},t) ,
\end{equation}
where $\alpha$ is an {\em arbitrary\/} constant.
Then, Eq.\ (\ref{b4}) reduces to
\begin{equation}
\label{b7}
\frac{\partial}{\partial t}\overline{f}-\frac{\partial}{\partial \overline{V}_i}
(a_{ij}\overline{V}_j+\alpha \overline{V}_i) \overline{f}
=Q[\overline{f},\overline{f}] .
\end{equation}
This equation can be interpreted as the one corresponding to USF in presence
of an external drag force $-m\alpha \overline{\bf V}$. Since the 
mapping of Eq.\ (\ref{b4}) onto Eq.\ (\ref{b7}) (and vice versa) is an exact 
property, we are free to choose the parameter $\alpha$ as we like.
For convenience, we choose $\alpha$ as a function of the shear rate
$a$ by the condition that the scaled temperature 
\beq
\overline{T}(t)=e^{-2\alpha t}T(t)=\frac{m}{3nk_B}\int d\overline{\bf V} \, 
\overline{ V}^2\overline{f}(\overline{\bf V},t)
\label{n1}
\eeq
reaches a constant value
in the long-time limit.
In the above equation, $m$ is the mass of a particle, $k_B$ is the Boltzmann 
constant, $n$ is the number density, and $T$ is the unscaled temperature.
With this choice of $\alpha$, the term $-m\alpha \overline{\bf V}$
plays the role of a {\em thermostat\/} force.
This kind of thermostat forces is usually employed in nonequilibrium molecular
dynamics simulations. \cite{H83,EM84,EM90}
Henceforth, we will adopt the point of view behind Eq.\ (\ref{b7}), i.e., all
the quantities will be understood to be scaled quantities, and we will drop
the bars for convenience. Taking second-degree moments in Eq.\ (\ref{b7}) we 
have
\begin{equation}
\label{b12}
\frac{\partial}{\partial t}P_{ij}+(a_{ik}P_{jk}+a_{jk}P_{ik})+2\alpha P_{ij}=
-\nu(P_{ij}-p\delta_{ij}),
\end{equation}
where
\begin{equation}
\label{b11}
P_{ij}=m\int d{\bf V}\,V_iV_jf
\end{equation}
is the pressure tensor, 
$p=nk_BT=\frac{1}{d}
\mathrm{Tr}\,{\sf P}$ is the hydrostatic pressure, and
 $\nu$ is an effective collision frequency defined as (see Appendix \ref{appA})
\begin{equation}
\label{b14}
\nu=n\frac{2d}{d-1}\int d\Omega\,K(\theta) \sin^2\frac{\theta}{2}\cos^2\frac{\theta}{2}
.
\end{equation}
It is convenient to choose $\nu^{-1}$ as the time unit and introduce the 
dimensionless quantities $t^*=\nu t$, $a^*=a/\nu$, and
$\alpha^*=\alpha/\nu$.
Henceforth, these reduced variables will be implicitly assumed and  
 the asterisks will be omitted.

{}From Eq.\ (\ref{b12}) it is easy to get the following closed differential 
equation
for the temperature:
\begin{equation}
\label{b19}                                                                
\left(\frac{\partial}{\partial t}+2\alpha\right)\left(\frac{\partial}{\partial 
t}+2\alpha+1\right)^2
T=\frac{2}{d}a^2T.
\end{equation}
So far, $\alpha$ is arbitrary.
Now, as said before, we choose $\alpha$ under the condition that the temperature
reaches a stationary value in the long-time limit. This implies that $\alpha(a)$ 
is the real root of the cubic equation
\begin{equation}
\label{b33}
a^2=d\alpha(1+2\alpha)^2,
\end{equation}
i.e.,
\begin{eqnarray}
\alpha(a)&=&\frac{2}{3}\sinh^2\left[\frac{1}{6}\cosh^{-1}\left(1+z\right)\right]
\nonumber\\
&=&\frac{1}{6}\left(1+z+\sqrt{2z+z^2}\right)^{1/3}
+\frac{1}{6}\left(1+z-\sqrt{2z+z^2}\right)^{1/3}-\frac{1}{3},
\label{b22}
\end{eqnarray}
{ where $z\equiv\frac{27}{d}a^2$.}
The stationary values of the elements of the pressure tensor are then easily
obtained from Eq.\ (\ref{b12}):
\beq
P_{xx}=p \frac{1+2d\alpha}{1+2\alpha},\quad P_{yy}=\cdots=P_{dd}=
\frac{p}{1+2\alpha}, \quad P_{xy}=-a\frac{p}{(1+2\alpha)^2}.
\label{b23}
\eeq
These quantities exhibit non-Newtonian effects: normal stress differences and a
nonlinear relationship between the shear stress and the shear rate.

While the second-degree moments are well defined for any value of the shear
rate $a$, moments of degree four and higher diverge if $a$ is larger than a
certain critical value $a_c$. \cite{SG95,SGBD93,MSG96,MSG97} This
suggests that the stationary solution to Eq.\ (\ref{b7}) presents a
high-velocity tail of the form $f({\bf V})\sim V^{-d-2-\sigma(a)}$, where
$\sigma(a)$ is a positive definite exponent that is a decreasing function of the
shear rate.
\section{A simple scattering model}
\label{sec3}
In the Monte Carlo simulations as well as in the theoretical analysis of higher-degree
moments presented in Ref.\ \onlinecite{MSG97} the scattering law was assumed to be
isotropic: $K(\theta)=\text{const}$. However, this choice for the collision rate
does not simplify the collision operator in a significant way { as} to 
allow one to 
 confirm theoretically the  high-velocity tail. Since the main 
 features
can be expected to be to some extent independent of the particular scattering
model  $K(\theta)$,  it is convenient to
consider the simplest choice lending itself to an analytic or semi-analytic
treatment. Here we consider the
following simplified scattering model:
\beq
K(\theta)=K_0 \lim_{\epsilon\to 0}\epsilon^{-d}\delta(\theta-\epsilon),
\label{c1}
\eeq
which corresponds to grazing collisions { of Maxwell molecules}. It is 
proved in Appendix \ref{appB}
that in this case the collision operator becomes
\beq
Q[f,f]\to \frac{\nu}{4d} 
\left[(d-1)\frac{\partial}{\partial V_{i}}V_{i}+\frac{dp\delta_{ij}-
P_{ij}}{mn}\frac{\partial}{\partial V_{i}}
\frac{\partial}{\partial
V_{j}}+{\cal L}^2\right]f({\bf
V}),
\label{c2}
\eeq
where
\beq
{\cal L}^2=\left({\bf V}\times\frac{\partial}{\partial {\bf
V}}\right)^2=\frac{\partial}{\partial V_{i}}
\left(V^2\delta_{ij}-V_{i}V_{j}\right)\frac{\partial}{\partial V_{j}}.
\label{c3}
\eeq
The right-hand side of (\ref{c2}) has the structure of a Fokker--Planck 
operator { with a velocity-dependent  diffusion term. 
The model (\ref{c2}) was derived (with $d=3$) in Section II.9 of 
Ref.~\onlinecite{C75} as the contribution of grazing collisions to the 
collision operator. Since here only grazing collisions are considered, the 
above contribution becomes the entire collision operator.}

\begin{figure}[t]
\centerline{\epsfig{file=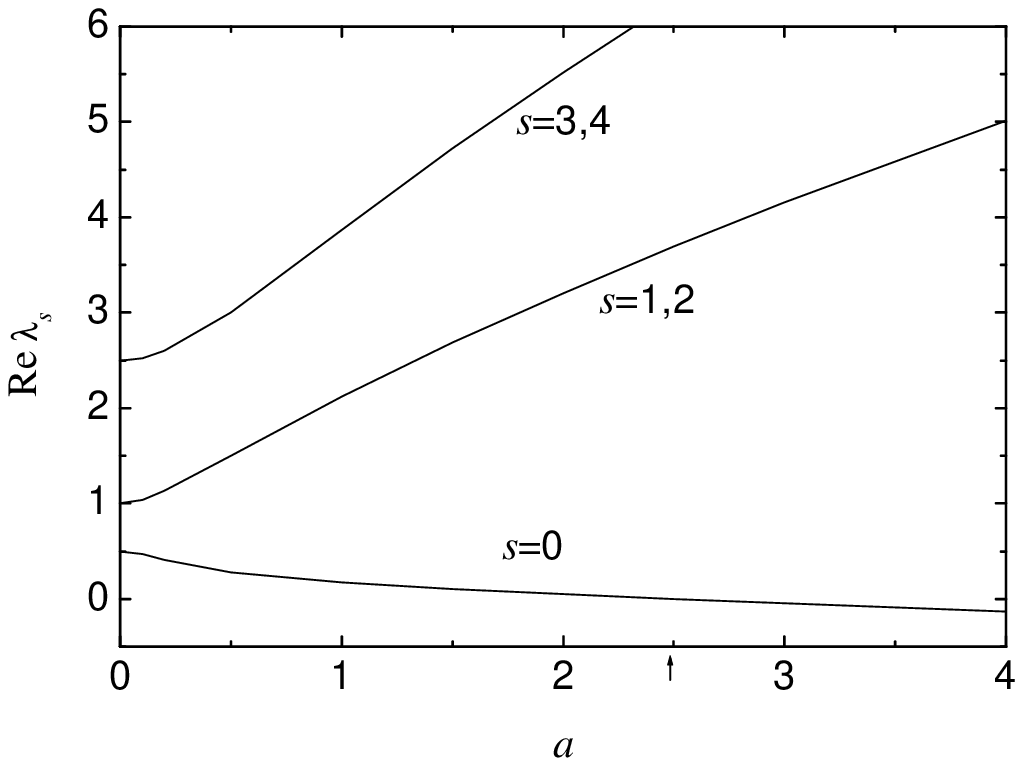,width=.7\textwidth}}
\caption{Real parts of the eigenvalues $\lambda_s$ associated with the
fourth-degree moments, as functions of the shear
rate. { The arrow indicates the location of the critical shear rate 
$a_c\simeq 2.48553$ beyond which $\lambda_0<0$ and, consequently, all the 
moments of degree four and higher diverge.}
\label{fig1}}
\end{figure}

Now, we restrict ourselves to the two-dimensional case ($d=2$), so that Eq.\
(\ref{b7}) becomes
\begin{equation}
\frac{\partial}{\partial t}{f}-aV_y\frac{\partial}{\partial {V}_x}f-\alpha
\frac{\partial}{\partial {V}_i}{V}_i{f}
=\frac{1}{8} 
\left(\frac{\partial}{\partial V_{i}}V_{i}+\frac{2p\delta_{ij}-
P_{ij}}{mn}\frac{\partial}{\partial V_{i}}
\frac{\partial}{\partial
V_{j}}+{\cal L}^2\right)f,
\label{c4}
\end{equation}
where we have already made $\nu=1$. Of course, by taking second-degree moments
we reobtain Eq.\ (\ref{b12}). Let us consider now the fourth-degree moments
\beq
M_s=\frac{1}{n}\int d{\bf V}\, V_x^{4-s}V_y^s f({\bf V}), \quad s=0,\ldots,4.
\label{c5}
\eeq
Multiplying both sides of Eq.\ (\ref{c4}) by $V_x^{4-s}V_y^s$ and integrating
over ${\bf V}$ we get
\beqa
\frac{\partial}{\partial t}M_s+(4-s)aM_{s+1}+4\alpha
M_s&=&-\frac{4+s(4-s)}{4}M_s+\frac{(4-s)(3-s)}{8}M_{s+2}
+\frac{s(s-1)}{8}M_{s-2}\nonumber\\
&&-\frac{s(4-s)}{4}P_1P_{s-1}+\frac{(4-s)(3-s)}{8}P_2P_{s}
+\frac{s(s-1)}{8}P_0P_{s-2},
\label{c6}
\eeqa
where $P_0\equiv P_{xx}/mn$, $P_1\equiv P_{xy}/mn$, and $P_2\equiv P_{yy}/mn$.
In matrix notation,
\beq
\frac{\partial}{\partial t} M_s=-\sum_{s'=0}^4\Lambda_{s s'}M_{s'}+N_s,
\label{c7}
\eeq
where 
\beq
\Lambda_{s 
s'}(a)=\left[4\alpha(a)+\frac{4+s(4-s)}{4}\right]\delta_{s',s}+(4-s)a
\delta_{s',s+1}-\frac{(4-s)(3-s)}{8}\delta_{s',s+2}-
\frac{s(s-1)}{8}\delta_{s',s-2},
\label{c8}
\eeq
and $N_s$ is a bilinear combination of second-degree moments.
The temporal evolution of the moments $M_s$ is governed by the eigenvalues
$\lambda_s(a)$ of the matrix $\Lambda_{s,s'}(a)$. At equilibrium ($a=0$) the
eigenvalues are $\lambda_0(0)=\frac{1}{2}$,  $\lambda_1(0)=\lambda_2(0)=1$, and
$\lambda_3(0)=\lambda_4(0)=\frac{5}{2}$. At finite shear rate, the degeneracy is
broken,  $(\lambda_1,\lambda_2)$ and $(\lambda_3,\lambda_4)$ becoming two
conjugate pairs. 
The trace of the matrix $\Lambda_{ss'}(a)$ is 
$\lambda_0(a)+2\text{Re}\lambda_1(a)+2\text{Re}\lambda_3(a)=\frac{15}{2}+20\alpha(a)$, 
which is a monotonically increasing function of the shear rate. On the other 
hand, as Fig.~\ref{fig1} shows, while the real parts of 
$\lambda_1$--$\lambda_4$ monotonically
increase with the shear rate, the only real eigenvalue $\lambda_0$ decreases and
eventually becomes negative for $a\geq a_c\simeq 2.48553$. This implies that for
$a\geq a_c$ the fourth-degree moments diverge in the long-time limit.
{ Of course, moments of a degree higher than four are also divergent for 
$a\geq a_c$. As said in the Introduction, the existence of a critical shear 
rate $a_c$ was first obtained analytically in the case of Maxwell molecules 
with regular collision rates, both for $d=3$\cite{SG95,SGBD93,MSG96} and 
$d=2$.\cite{MSG97,MGS97} For instance, $a_c\simeq 5.847$\cite{MSG97} in the 
case of two-dimensional Maxwell molecules with isotropic scattering. Thus, 
the singular behavior of the fourth-degree moments $M_s$ is not an artifact 
of the scattering model (\ref{c1}).}

{ The divergence of $M_s$ for $a\geq a_c$}  is a
strong indication that the velocity distribution function reaches a steady state
form that has an algebraic decay $f({\bf V})\sim V^{-4-\sigma(a)}$ with
$\sigma(a)\leq
2$. Conversely, for $a<a_c$ one can expect that $\sigma(a) >2$, so that the fourth-degree
moments converge, but  moments of a higher degree  $2k+2$ will diverge whenever the
shear rate is such that $\sigma(a)\leq 2k$.  
While this scenario has been supported by computer simulations,
\cite{MSG97,MGS97} we
are not aware of any derivation of this high-velocity tail from 
the Boltzmann
equation, { prior to the one presented in the next Section.}
\section{High-velocity tail}
\label{sec4}

{ {}From Eq.~(\ref{c4}) we get the following equation for the steady 
state distribution function:
\begin{equation}
\left[8aV_y\frac{\partial}{\partial {V}_x}+(1+8\alpha)
\frac{\partial}{\partial {V}_i}{V}_i+{\cal L}^2\right]f=
\frac{
P_{ij}-2p\delta_{ij}}{mn}\frac{\partial}{\partial V_{i}}
\frac{\partial}{\partial V_{j}}f.
\label{d4}
\end{equation}
This is still a complicated equation to solve for all velocities. However, 
we are interested here in the solution for large velocities. In that 
domain, the right-hand side of Eq.~(\ref{d4}) is of order $V^{-2}$ 
relative to the term ${\cal L}^2 f$ [cf.\ Eq.~(\ref{c3})], so it can be 
neglected.
By the arguments given at the end of the previous Section, we look 
for self-consistent solutions with the asymptotic behavior
\beq
f({\bf V};a)\sim V^{-4-\sigma(a)}\Phi(\varphi;a)
\label{d1}
\eeq
for large velocities. In Eq.\ (\ref{d1}) $\varphi$ is the polar angle in
velocity space, i.e.,
\beq
V_x=V\cos \varphi,\quad V_y=V\sin \varphi,
\label{d2}
\eeq
$\Phi$ is a function measuring the degree of anisotropy of the high-velocity
distribution function, and $\sigma$ is the exponent characterizing the algebraic
decay. These two quantities must be determined self-consistently. According 
to Eq.~(\ref{d1}), the left-hand side of Eq.~(\ref{d4}) is proportional to 
$V^{-4-\sigma}$, while the right-hand side is proportional to 
$V^{-6-\sigma}$. Consequently, the latter must be neglected against the 
former in the limit of large velocities, as said before.}

In polar coordinates, { the operators on the left-hand side of 
Eq.~(\ref{d4}) become}
\beq
V_y\frac{\partial}{\partial {V}_x}=\cos\varphi\sin\varphi
V\frac{\partial}{\partial {V}}-\sin^2\varphi \frac{\partial}{\partial \varphi},
\label{c5.2}
\eeq
\beq
\frac{\partial}{\partial {V}_i}{V}_i =2+V\frac{\partial}{\partial {V}},
\label{c6.2}
\eeq
\beq
{\cal L}^2=\frac{\partial^2}{\partial \varphi^2}.
\label{c7.2}
\eeq
Thus, when Eq.\ (\ref{d1}) is inserted into Eq.\ (\ref{d4}) one gets the
following linear second-order ordinary differential equation
\beq
\Phi''(\varphi)-p\left[1-\cos(2\varphi)\right]\Phi'(\varphi)-\left[\beta+2q\sin(2\varphi)\right]
\Phi(\varphi)=0, 
\label{d3}
\eeq
where $p\equiv 4a$, $\beta\equiv (2+\sigma)(1+8\alpha)$, and
$q\equiv 2a(4+\sigma)$. 
Equation (\ref{d3}) is a generalization of Mathieu's equation \cite{AS,GR} and reduces
to it if $p=0$. However, since in our case $p\neq 0$, we have to deal with Eq.\
(\ref{d3}) rather than with  Mathieu's well-known equation. The symmetry property
$f({\bf V})=f(-{\bf V})$ implies the periodicity condition
$\Phi(\varphi)=\Phi(\varphi+\pi)$. {}From a mathematical point of view, 
Eq.\ (\ref{d3}) supports those periodic
solutions provided that the parameter $\beta$ takes a  
characteristic value $\beta(p,q)$. Since the shear rate $a$ and the scaling
coefficient
$\alpha$ are related through Eqs.\ (\ref{b33}) and (\ref{b22}), only two of the
three parameters $\beta$, $p$, and $q$ are independent. Thus, the characteristic
value $\beta(p,q)$ translates into  $\sigma(a)$. In case there are
multiple solutions for a given $a$, the relevant
solution is the one related to the dominant tail, which corresponds to the
 smallest value of $\sigma$.
In the subsequent { mathematical} analysis we consider $\beta$, $p$, and 
$q$ as independent parameters and
only at the end we will take into account that  in our physical problem  $\beta=
(2+\sigma)(1+8\alpha)$, $p= 4a$, and $q= 2a(4+\sigma)$.

In order to find the periodic solutions to Eq.\ (\ref{d3}) and the corresponding
characteristic values, we write
\beq
\Phi(\varphi)=\sum_{m=-\infty}^\infty C_m e^{2im\p}, \quad C_{-m}=C_m^*.
\label{d4.2}
\eeq
If the above is substituted into Eq.\ (\ref{d3}), one obtains the recurrence
relations
\beq
\mu_{m+1}C_{m+1}+\gamma_m C_m-\mu_{-m+1}C_{m-1}=0,
\label{d5}
\eeq
where
\beq
\mu_m\equiv q- mp,\quad \gamma_m\equiv 2mp-i(\beta+4m^2).
\label{d6}
\eeq
In particular, setting $m=0$,
\beq
C_0=-i\frac{q-p}{\beta}\left(C_1-C_{-1}\right)=2\frac{q-p}{\beta}\text{Im} C_1.
\label{d7}
\eeq
Therefore, by applying Eq.\ (\ref{d5}) to $m\geq 1$, all the coefficients 
$C_m$,
$m\geq 2$, can be obtained recursively from $C_1$. The coefficients $C_m$ with
$m\leq -2$ are then given by the symmetry relation $C_{-m}=C_m^*$. The
characteristic values arise from the convergence condition
\beq
\lim_{m\to \infty} |D_m|<1, \quad D_m\equiv \frac{C_{m+1}}{C_m}.
\label{d8}
\eeq
The recurrence relation for the ratios $D_m$ is
\beq
\mu_{m+1}D_{m}+\gamma_m-\mu_{-m+1}D_{m-1}^{-1}=0.
\label{d9}
\eeq
Thus, all the coefficients $D_m$ with $m\geq 1$ are obtained from $D_0$, the
latter being subject to the compatibility condition
\beq
\text{Im} D_0=\frac{\beta}{2(q-p)},
\label{d10}
\eeq
as follows from Eq.\ (\ref{d7}). Equation (\ref{d9}) shows that there are two
possible asymptotic behaviors of $D_m$ for large $m$. Either
\beq
D_m\approx -\frac{4i}{p} m
\label{d11}
\eeq
or
\beq
D_m\approx -\frac{p}{4i} m^{-1}.
\label{d12}
\eeq
The latter is the only one consistent with the convergence condition (\ref{d8}).
This allows us to use (\ref{d9}) to develop $D_m$ as the continued fraction
\begin{mathletters}
\beqa
D_m&=&\frac{\mu_{-m}}{\gamma_{m+1}+\mu_{m+2}D_{m+1}}
\label{d13a}
\\
&=&\frac{\mu_{-m}}{\gamma_{m+1}+}\frac{\mu_{m+2}\mu_{-m-1}}{\gamma_{m+2}+}
\frac{\mu_{m+3}\mu_{-m-2}}{\gamma_{m+3}+}\cdots.
\label{d13b}
\eeqa
\end{mathletters}
Finally, the characteristic values are obtained from the compatibility condition
(\ref{d10}), namely
\beq
\text{Im}\left(\frac{\mu_{0}}{\gamma_{1}+}\frac{\mu_{2}\mu_{-1}}{\gamma_{2}+}
\frac{\mu_{3}\mu_{-2}}{\gamma_{3}+}\cdots\right)=\frac{\beta}{2(q-p)}.
\label{d14}
\eeq
Once the relationship $\beta(p,q)$ is found, the ratios $D_m$ are obtained from
Eq.\ (\ref{d13b}) or, equivalently, from Eq.\ (\ref{d13b}) for $m=0$ and from Eq.\
(\ref{d9}) for $m\geq 1$. The angular distribution function is then given by
Eq.\ (\ref{d4.2}) with
\beq
C_m=\frac{1}{2\pi}\prod_{m'=0}^{m-1}D_{m'},\quad m\geq 1,
\label{d14.2}
\eeq
where the particular value $C_0=1/2\pi$ has been chosen to verify the normalization condition
\beq
\int_0^{2\pi} d\p\, \Phi(\p)=1.
\label{d15}
\eeq
Since Eqs.\ (\ref{d13b}) and (\ref{d14}) involve infinite continued fractions,
in practice we proceed by setting $D_N=-(p/4i)N^{-1}$ for a certain large value of $N$
and then using (\ref{d13a}) to get $D_m$ for $m\leq N-1$ up to $m=0$. We have
typically taken $N\sim 10^3$ and have checked that the results are rather
insensitive to a further increase of $N$.

Before going back to our original physical problem and take into account the
expressions of $\beta$, $p$, and $q$ in terms of $a$ and $\sigma$, it is
worthwhile exploring the following limiting cases. 
\begin{enumerate}
\item[(i)]
\label{i}
Consider first that the ratio $q/p$ is
equal to an integer number, $q/p=k+2$. This means that $\mu_{k+2}=0$, so that
Eq.\ (\ref{d14}) becomes
\beq
\text{Im}\left(\frac{\mu_{0}}{\gamma_{1}+}\frac{\mu_{2}\mu_{-1}}{\gamma_{2}+}
\frac{\mu_{3}\mu_{-2}}{\gamma_{3}+}\cdots \frac{\mu_{k+1}\mu_{-k}}{\gamma_{k+1}}
\right)=\frac{\beta}{2(k+1)p},
\label{d16}
\eeq
which gives rise to an algebraic equation of degree $2k+3$ for $\beta$. For
instance,
\beq
\beta^3+8\beta^2+16\beta-16 p^2=0,
\label{d16.1}
\eeq
\beq
\beta^5+40\beta^4+528\beta^3+16(160-21p^2)\beta^2+256(16-15p^2)\beta-12288p^2=0,
\label{d16.2}
\eeq
for $k=0$ and $k=1$, respectively.
Next, Eq.\ (\ref{d13a}) gives $D_m$ for $m\leq k$ in a closed form, while the
infinite continued fraction (\ref{d13b}) must be used for $m\geq k+1$. For
instance, in the case $k=0$,
\beq
D_1=\frac{4p}{4p-(16+\beta)i},\quad D_0=3p
\frac{4p-(16+\beta)i}{12p^2-(4+\beta)(16+\beta)-6p(8+\beta)i}.
\label{d16.3}
\eeq
\item[(ii)]
\label{ii}
We now consider  the limit $q\to \infty$ with
$p=\text{finite}$. In that case, as
verified later by consistency, the characteristic value scales as
$\beta\to q\beta_1$, so that $\mu_m\to q$ and $\gamma_m\to -i q\beta_1$.
Consequently, Eq.\ (\ref{d9}) becomes
\beq
D_m -i\beta_1-D_{m-1}^{-1}=0.
\label{d17}
\eeq
The solution consistent with the convergence of the series and with Eq.\
(\ref{d10}) is simply $D_m=e^{-2i\p_0}$, where $\p_0$ is an angle defined by
\beq
 \sin 2\p_0=-\frac{\beta_1}{2}.
\label{d18}
\eeq
This implies that $0\leq \beta_1\leq 2$. The function $\Phi$ is 
\beqa
\Phi(\p)&=&\frac{1}{2\pi}\sum_{m=-\infty}^\infty e^{2im(\p-\p_0)}\nonumber\\
&=&\frac{1}{2}\left[\delta(\p-\p_0)+\delta(\p-\p_0+\pi)\right].
\label{d19}
\eeqa
\item[(iii)]
\label{iii}
As a third limiting case, let us consider $q\to \infty$, $p\to \infty$,
$q/p=\text{finite}$. It is then expected that $\beta/q\to 0$, so that
$\gamma_m\to 2mp$. The consistent solution to Eq.\ (\ref{d9}) is $D_m=1$, which
corresponds to
\beq
\Phi(\p)=\frac{1}{2}\left[\delta(\p)+\delta(\p-\pi)\right].
\label{d20}
\eeq
Numerical analysis shows that in this limit the characteristic value $\beta$
scales as $\beta=(\beta_2 p^2)^{1/3}$, where $\beta_2$ depends on the ratio
$q/p$. { In the special case of $q/p=\text{integer}$, the values of 
$\beta_2$ can be obtained from the algebraic equation
(\ref{d16}).} In particular, we
get $\beta_2=16$, $\beta_2=336$, and $\beta_2=1008\pm 96\sqrt{79}$ for
$q/p=2$, $q/p=3$, and $q/p=4$, respectively.
\end{enumerate}

Now we apply the above analysis to the physical values
$\beta=(2+\sigma)(1+8\alpha)$, $p= 4a$, and $q= 2a(4+\sigma)$, where $a$ and
$\alpha$ are related by Eqs.\ (\ref{b33}) and (\ref{b22}). The case (i) above,
namely
$q/p=k+2$ with $k$ integer, corresponds to $\sigma=2k$, so that the solution
to Eq.\ (\ref{d16}) gives the critical value of the shear rate beyond which
all the moments of a degree equal to or larger than $2k+2$  diverge. For
$k=0$, Eq.\ (\ref{d16.1}) gives rise to a cubic equation in $\alpha$ with no positive real
root. This is expected since, as seen in Sec.\ \ref{sec2}, the second-degree
moments converge. For
$k=1$, Eq.\ (\ref{d16.2}) gives rise to the same quintic equation as obtained
from $\det \Lambda_{ss'}=0$, where the matrix $\Lambda_{ss'}$ is given by Eq.\
(\ref{c8}). The solution to this equation is $\alpha\simeq 0.61796$, which
corresponds to $a\simeq 2.48553$, i.e., the critical value $a_c$ found in
Sec.\ \ref{sec2} from the time evolution of the fourth-degree moments.
In a similar way, we find the values $a\simeq 1.11175$, $a\simeq 0.87611$,
$a\simeq 0.77052$, and $a\simeq 0.70842$ for $\sigma=4$, $\sigma=6$, $\sigma=8$,
and $\sigma=10$, respectively. It is interesting to note that for $\sigma=8$
and $\sigma =10$ there exists a larger second  solution ($a\simeq 52.5$ and
$a\simeq 7.3$, respectively). This indicates that, in addition to
the smallest eigenvalue, another of the eigenvalues governing
the time evolution of the moments of degree 10 and 12, respectively, becomes
negative at those large shear rates. This property has been observed before in the
case of more realistic scattering laws. \cite{MSG97} Of course, whenever there
are different values of $a$ consistent with a given value of $\sigma$, the
{ actual} function $\sigma(a)$  corresponds to the smallest value of $a$.
Equivalently, if
the characteristic equation (\ref{d14}) gives more than a value of $\sigma$
for a given shear rate, the function $\sigma(a)$ is defined by the smallest
value.

Now we consider the case (ii) above, which corresponds to the situation
$\sigma\to\infty$. For which value of the shear rate does this happen?
Mathematically speaking, there are an infinite number of solutions, namely the
solutions to $(1+8\alpha)/2a=\beta_1$ with $0\leq\beta_1\leq 2$. Taking into
account Eq.\ (\ref{b33}), this is equivalent to the cubic equation
\beq
8\beta_1^3 a^3+4(5\beta_1^2-16)a^2+6\beta_1 a-9=0.
\label{d21}
\eeq
The real solution to this equation is a monotonically decreasing function of $\beta_1$.
Therefore, the physical value of the shear rate at which $\sigma\to \infty$
corresponds to $\beta_1=2$. This defines the threshold shear rate
$a_{\text{th}}=(\sqrt{57}/96+67/864)^{1/3}-(\sqrt{57}/96-67/864)^{1/3}-1/12\simeq
0.352047$ beyond which { the high-velocity tail has the form 
(\ref{d1}).}
Conversely, if $a\leq a_{\text{th}}$, the decay of the velocity distribution
function is more rapid than any algebraic tail and {\em all\/} the velocity
moments are convergent. This is in contrast with the results obtained from
realistic scattering laws, which suggest that $a_{\text{th}}\to 0$ in those
cases.\cite{MSG96,MSG97}
Since $\beta_1=2$, Eq.\ (\ref{d18}) yields $\p_0=3\pi/4$ and Eq.\ (\ref{d19})
becomes
\beq
\lim_{a\to a_{\text{th}}}\Phi(\p;a)=
\frac{1}{2}\left[\delta\left(\p-\frac{3}{4}\pi\right)+\delta\left(\p-\frac{7}{4}\pi\right)\right],
\quad \lim_{a\to a_{\text{th}}}\sigma(a)=\infty.
\label{d22}
\eeq

Next, we investigate the limit of large shear rates, $a\to\infty$. In that
case, $\alpha \approx a^{2/3}/2$, on account of Eq.\ (\ref{b33}). The
question now is, what is the corresponding value $\sigma(\infty)\equiv
\sigma_{\text{min}}$? Computer simulations for Maxwell molecules interacting
with an isotropic scattering law suggest that $\sigma_{\text{min}}=0$.
However, this is not the case with the scattering law (\ref{c1}). According
to the case (iii) above, the value of $\sigma_{\text{min}}$ must be such that $\beta_2=
4(2+\sigma_{\text{min}})^3$ for $q/p=2+\sigma_{\text{min}}/2$. A numerical calculation gives
$\sigma\simeq 1.252$, i.e., $\beta_2\simeq 137.6$ at $q/p=2.626$, which is
consistent with the fact that $\beta_2=16$ and $\beta_2=336$ at $q/p=2$ and
$q/p=3$, respectively. The minimum value $\sigma_{\text{min}}$ means that, no
matter how large the shear rate is, not only the second-degree moments are
finite, but so are moments of the form $\langle V^k\rangle$ with $k\leq
2+\sigma_{\text{min}}\simeq 3.252$. On the other hand, the high-velocity
distribution function becomes strongly anisotropic for large shear rates
since, according to Eq.\ (\ref{d20}),
\beq
\lim_{a\to\infty}\Phi(\p;a)=\frac{1}{2}\left[\delta(\p)+\delta(\p-\pi)\right],
\quad \lim_{a\to \infty}\sigma(a)=\sigma_{\text{min}}.
\label{d23}
\eeq

\begin{figure}[t]
\centerline{\epsfig{file=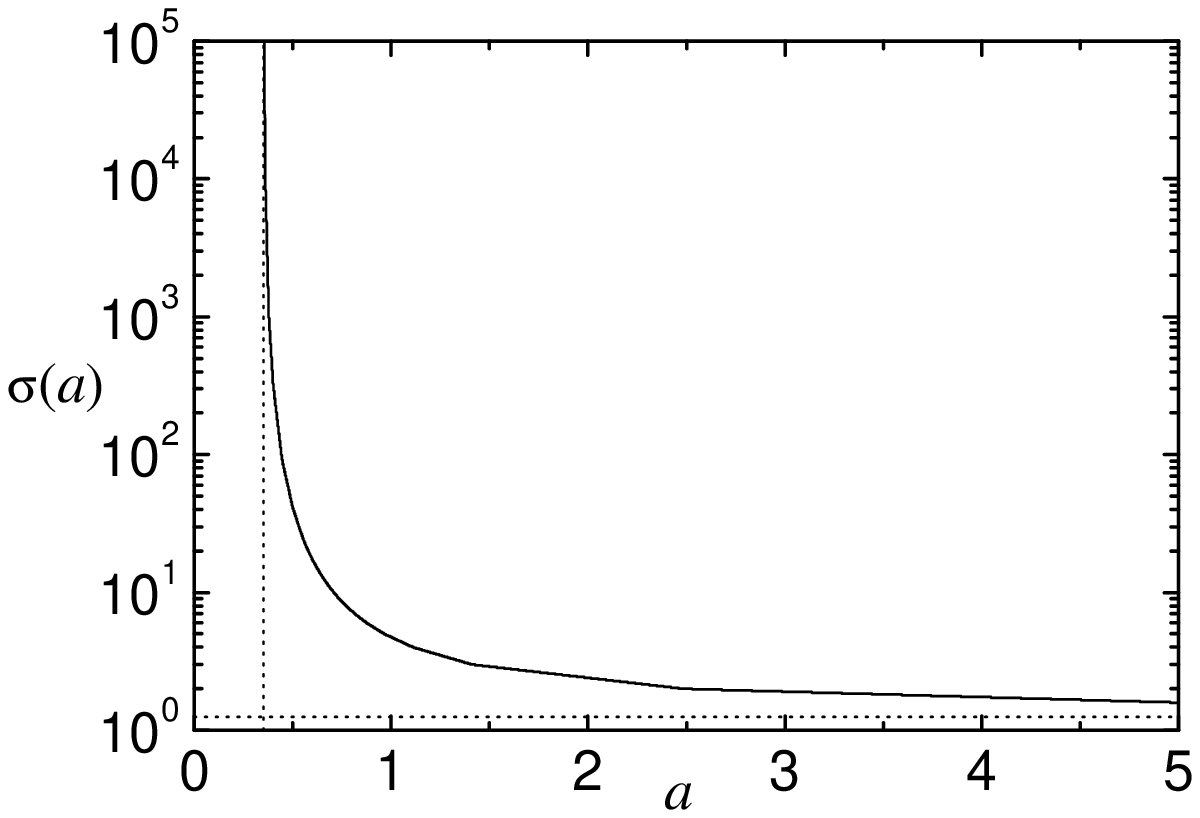,width=.7\textwidth}}
\caption{Plot of the exponent $\sigma$ as a function of the shear rate $a$. 
The vertical and horizontal dotted lines indicate the locations of 
$a_{\text{th}}\simeq 0.3520$ and $\sigma_{\text{min}}\simeq 1.252$, 
respectively.
\label{fig2}}
\end{figure}

\begin{figure}[t]
\centerline{\epsfig{file=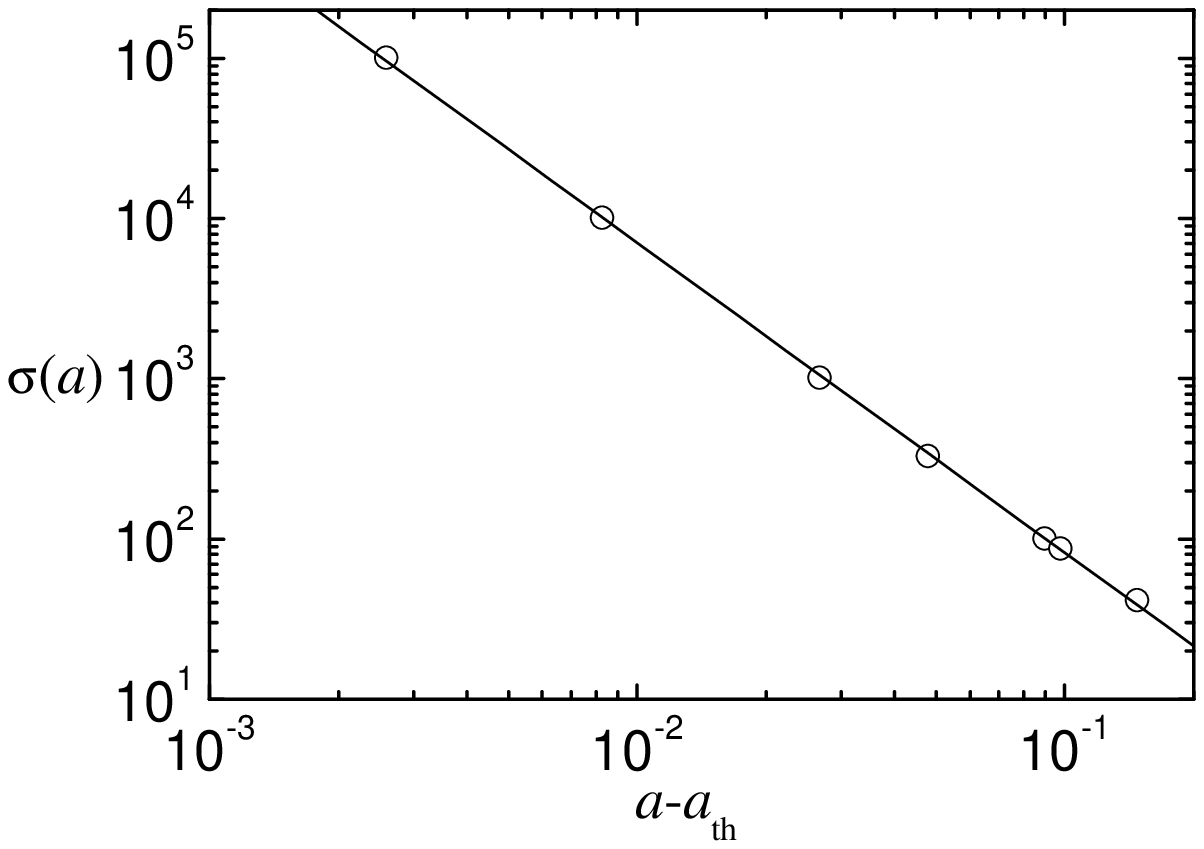,width=.7\textwidth}}
\caption{Log-log plot of  $\sigma$ versus $a-a_{\text{th}}$. The solid line
is a linear fit.
\label{fig3}}
\end{figure}

\begin{figure}[t]
\centerline{\epsfig{file=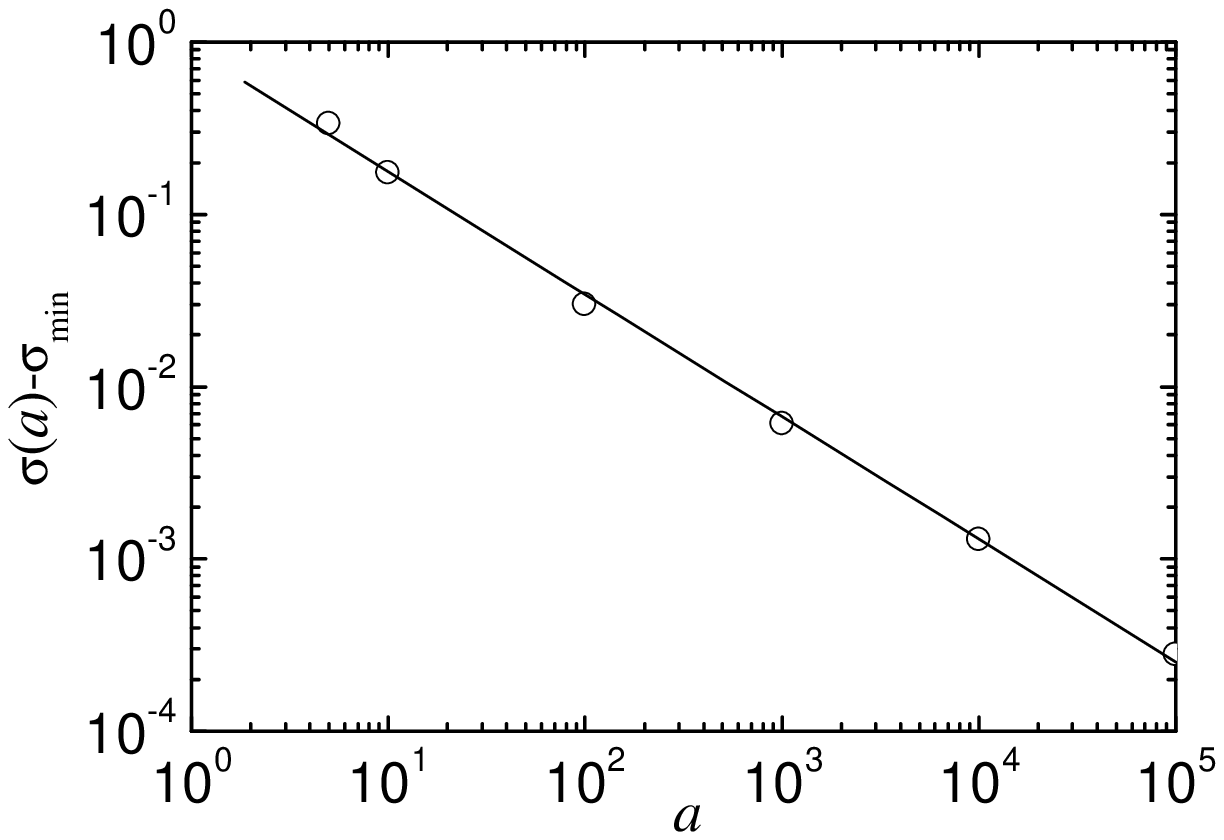,width=.7\textwidth}}
\caption{Log-log plot of  $\sigma-\sigma_{\text{min}}$ versus $a$. The solid line
is a linear fit.
\label{fig4}}
\end{figure}

The complete dependence of $\sigma$ on the shear rate is shown in Fig.\
\ref{fig2}, where the dotted lines indicate the locations of $a_{\text{th}}$
and $\sigma_{\text{min}}$. The region near $a=a_{\text{th}}$  is shown in 
Fig.~\ref{fig3}. The results  can be well fitted to the { power law}
$\sigma\approx e^{-0.02}(a-a_{\text{th}})^{-1.93}$. The behavior for large
shear rates is shown in Fig.~\ref{fig4}, where the results can be fitted to
$\sigma-\sigma_{\text{min}}\approx e^{-0.04}a^{-0.71}$.

\begin{figure}[t]
\centerline{\epsfig{file=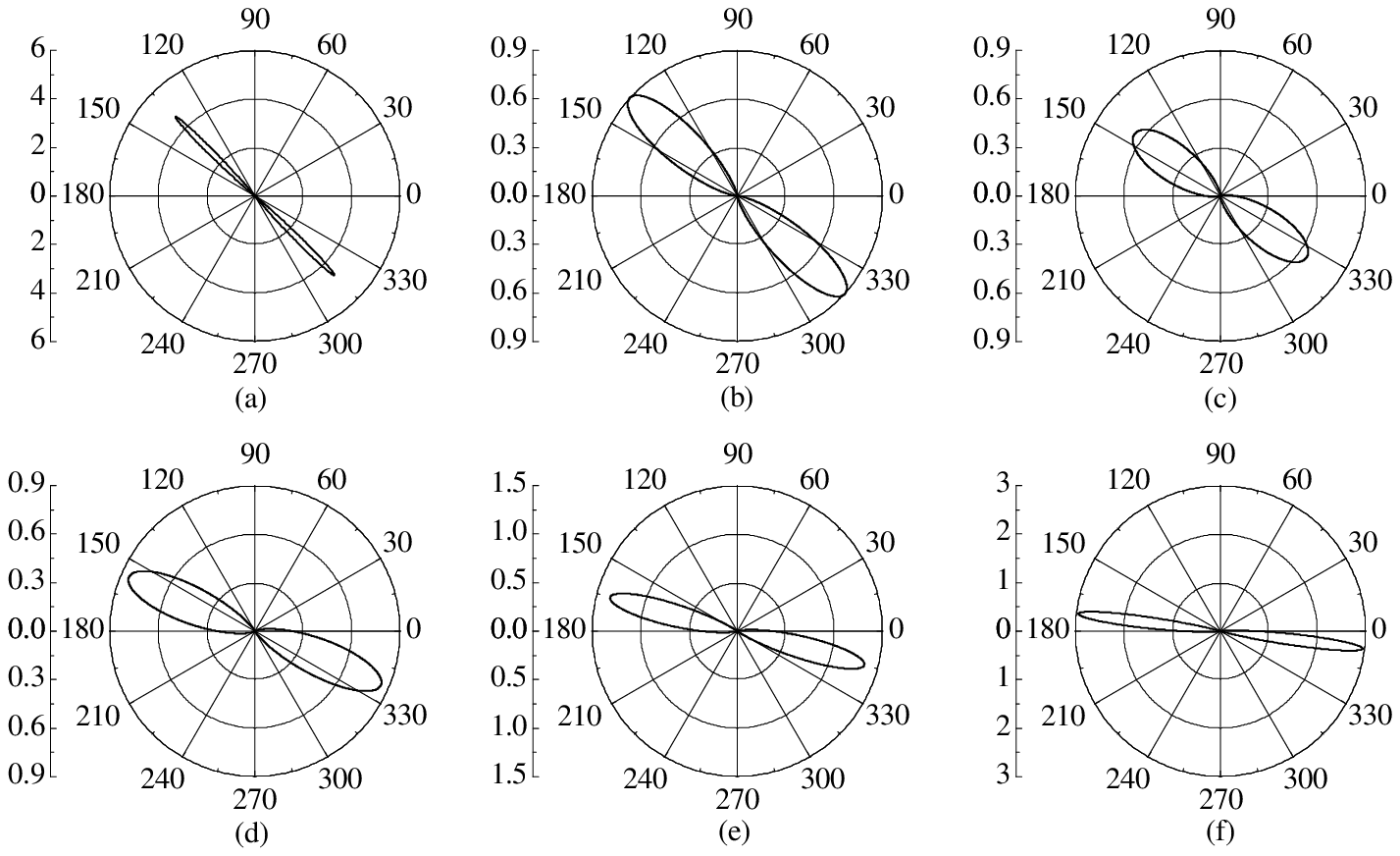,width=.7\textwidth}}
\caption{Polar diagram of $\Phi(\p)$ for (a) $a=0.354652$ ($\sigma=10^5$),
(b) $a=0.442229$ ($\sigma=10^2$), (c) $a=0.708415$ ($\sigma=10$), (d)
$a=2.485530$ ($\sigma=2$), (e) $a=10$ ($\sigma=1.427204$), and (f) $a=10^2$
($\sigma=1.282082$). { Case (a) corresponds to a shear rate slightly 
larger than the threshold value $a_{\text{th}}\simeq 0.3520$, while case (f) 
corresponds to an exponent $\sigma$ slightly larger than the minimum value 
$\sigma_{\text{min}}\simeq 1.252$. Note that the scales in cases (a), (e), 
and (f) are different from the scales in cases (b)--(d).}
\label{fig5}}
\end{figure}

In addition to the function $\sigma(a)$, the solution to the problem gives
the angular distribution $\Phi(\p;a)$ for any shear rate $a>a_{\text{th}}$. The
results indicate that in the high-velocity domain the distribution is
highly anisotropic, with the angular distribution sharply concentrated around
a preferred orientation angle $\widetilde{\p}(a)$ that rotates from
$\widetilde{\p}=3\pi/4,7\p/4$ when $a\to a_{\text{th}}$, Eq.\ (\ref{d22}), to
$\widetilde{\p}=0,\pi$ when $a\to a_\infty$, Eq.\ (\ref{d23}). This transition
is illustrated in Fig.\ \ref{fig5}.
{ The anisotropic behavior observed in Fig.~\ref{fig5} is consistent with 
the one shown in Fig.~8 of Ref.~\onlinecite{MSG97} in the case of computer 
simulations of Maxwell molecules with a constant collision rate.}
\section{Concluding remarks}
\label{sec5}
In this paper we have shown that the stationary solution of the (thermostatted)
Boltzmann--Fokker--Planck equation for a two-dimensional gas of Maxwell 
molecules under uniform
shear flow exhibits an algebraic high-velocity tail of the form given by Eq.\
(\ref{d1}), where $\sigma(a)$ is a decreasing function of the shear rate $a$.
As a consequence, all the velocity moments of a degree equal to or larger than
$2+\sigma(a)$ diverge. The word ``thermostat'' does not mean in the present
context that an external drag force is necessarily applied on the system;
its effect for Maxwell molecules is equivalent to a rescaling of the
velocities with respect to the thermal velocity and this is the point of view
adopted in this paper. 
The restriction to  an idealized scattering
model with grazing collisions, Eq.\ (\ref{c1}), allows us to replace the
complicated structure of the Boltzmann collision operator by a much more
manageable Fokker--Planck operator, { Eq.~(\ref{c2}).} This operator is 
formally linear in
velocity space, but the original bilinear nature of the collision operator 
appears via the dependence of
some of the coefficients on the distribution function through the pressure
tensor. On the other hand, those nonlinear terms are negligible in the limit
of large velocities and thus one arrives at a linear second-order ordinary differential
equation for $\Phi$, Eq.\ (\ref{d3}). Despite the linearity of the problem
posed by Eq.\ (\ref{d3}), it applies to states arbitrarily far from
equilibrium. As a matter of fact, the nonlinear nature of the underlying state is still
present through
the dependence of the ``thermostat'' (or scaling) parameter $\alpha$ on the
shear rate $a$, Eq.\ (\ref{b22}). The structure of Eq.\ (\ref{d3}) is
reminiscent of Mathieu's equation; in fact the latter is obtained by formally
setting the parameter $p=0$, although this is not { possible in} 
the physical problem at hand. The symmetry property $f({\bf V})=f(-{\bf 
V})$ translates into the
periodicity condition $\Phi(\varphi)=\Phi(\varphi+\pi)$. 
{ Enforcements of this condition on the solutions to  Eq.~(\ref{d3}) 
allows us to} obtain $\sigma(a)$ from 
the
solution of the continued fraction representation (\ref{d14}), the latter
becoming an algebraic equation when $\sigma=2,4,6,\ldots$.
The results show that $\sigma(a)$ is lower bounded by the value
$\sigma_{\text{min}}\simeq 1.252$ (which is asymptotically reached in the
limit $a\to\infty$) and goes to infinity when the shear rate approaches
the threshold value $a_{\text{th}}\simeq 0.352$ (in units of the collision
frequency). The first property implies 
that all the moments of degree equal to or smaller than $2+\sigma_{\text{min}}\simeq 3.252$ 
are finite
in the steady state, regardless of how large the shear rate is. As a
consequence of the second property, there exists a window of shear rates
$0\leq a\leq  a_{\text{th}}$ where the high-velocity population decays more
rapidly than algebraically { (e.g., as a stretched exponential),} all the 
moments being finite. The lobular shape
of the orientation distribution $\Phi(\varphi)$ [cf.\  Fig.\ \ref{fig5}]
shows that the high-velocity population is highly anisotropic, a feature already
observed in Monte Carlo simulations  for the isotropic scattering
model.\cite{MSG97}
Most of the particles having a large (peculiar) velocity ${\bf V}$ move along a narrow
bunch of directions  around a preferred direction characterized by the
(equivalent) angles
$\widetilde{\p}(a)$ and $\widetilde{\p}(a)+\pi$. The angle
$\widetilde{\p}(a)$ rotates counter-clockwise from $\widetilde{\p}=3\pi/4$
(i.e., $V_y/V_x=-1$) at
$a=a_{\text{th}}$ to $\widetilde{\p}=\pi$ (i.e., $V_y/V_x\to 0^-$) 
in the limit $a\to\infty$. The
width of the angular distribution is zero in both limits, being maximum at a
shear rate $a\approx 1$. 

It is worth remarking that, despite the simplicity and artificiality of the
scattering model considered in this paper, it succeeds in capturing the most 
relevant features that were expected on the basis of
 Monte Carlo simulations 
and   moment
method results in the case of  more realistic scattering 
{ laws:}\cite{MSG96,MSG97} the existence of a
high-velocity tail with a monotonically decreasing exponent $\sigma(a)$ and
with a strongly anisotropic orientation distribution $\Phi(\p;a)$. 
{ Thus, the asymptotic behavior (\ref{d1}) is not an artifact of the 
specific model (\ref{c2}), but a general property of Maxwell molecules under 
uniform shear flow. Some other features, however, are possibly peculiar of 
the singular scattering law (\ref{c1}). For instance,}
the results of Ref.\ \onlinecite{MSG97} for the isotropic
scattering model seem to indicate that $\sigma(a)$ only diverges when $a\to 0$ (i.e.,
$a_{\text{th}}=0$) and that $\lim_{a\to \infty}\sigma(a)=0$ (i.e.,
$\sigma_{\text{min}}=0$). It is possible that the extreme anisotropy of the
scattering model (\ref{c1}) makes the high-velocity tail phenomenon to be
milder than in the general case, thus yielding non-zero values for
$a_{\text{th}}$ and $\sigma_{\text{min}}$.
On a different vein, it must be said that { since the 
problem reduces to a
linear equation, we have not determined} either the 
coefficient measuring
the amplitude of the tail (\ref{d1}) or the order of magnitude of the
characteristic velocity $c$, such that (\ref{d1}) applies when $V\gg c$. Monte
Carlo simulations \cite{MSG97} show that the amplitude is a decreasing function
of the shear rate, while  $c$ is rather insensitive to the shear rate. { 
The value of $c$ seems to be} of the order of the thermal velocity
$v_0=\sqrt{2k_BT/m}$, so that the asymptotic behavior (\ref{d1}) is 
reached in practice for $V\gtrsim 10 v_0$. The theoretical confirmation of both
properties would require the analysis of the full kinetic equation (\ref{d4})
and this is beyond the scope of this work.

{ It must be remarked that in this paper we have considered the 
replacement (\ref{c2}) as a collision model by itself. On the other hand, it 
is known that the right-hand side of (\ref{c2}) actually represents the 
contribution to the true collision operator associated with grazing 
collisions.\cite{C75} Therefore, it is tempting to conjecture that, in 
general,  the 
high-velocity tail is produced by grazing collisions of high-velocity 
particles. Since the expansion leading to (\ref{c2}) is non-uniform for high 
velocities,\cite{C75} the analysis of this paper does not provide a rigorous 
proof of the above conjecture but, at most, a strong indication of it.}

The results presented in this paper, along with those of Refs.\ 
\onlinecite{SG95,SGBD93,MSG96,MSG97,MGS97}, show that the population of high 
energy levels in the uniform shear flow, and possibly in many other 
nonequilibrium states, is far greater than in the corresponding equilibrium 
state described by the Maxwell--Boltzmann distribution function. 
{ Among other examples of nonequilibrium states exhibiting algebraic 
high-velocity tails we can mention the viscous longitudinal flow of Maxwell 
molecules\cite{S00} and the homogeneous cooling state of inelastic Maxwell 
molecules.\cite{EB01} As a consequence of these nonequilibrium 
overpopulation effects,}
 many more particles { than at equilibrium} can be available for 
surmounting the energy barriers of chemical and nuclear reactions which, 
consequently, would proceed faster. We also guess that fluctuations are 
enhanced with respect to the equilibrium ones, but the analysis of this 
expectation would require the treatment of the fluctuating Boltzmann 
equation.\cite{C80} This is an interesting avenue for further theoretical 
work.

\acknowledgments
L.A. and A.S.
acknowledge partial support from the Ministerio de Ciencia y Tecnolog\'{\i}a
 (Spain) through grant No.\ BFM2001-0718. This work was initiated while 
 A.S. and A.V.B. were 
visiting the Department of Aeronautics and Astronautics, Graduate School of 
Engineering, Kyoto University, as visiting fellows of the Japan Society for 
the Promotion of Sciences. They are grateful to Profs.\ Y. Sone and K. Aoki 
for their kind hospitality. 
\appendix
\section{Derivation of Eq.\ \lowercase{(\protect\ref{b14})}}
\label{appA}
The Boltzmann collision operator for Maxwell molecules is
\beq
Q[{\bf V}_1|f,f]=\int d{\bf V}_2\int d\Omega\, K(\theta)\left[f({\bf V}_1')
f({\bf V}_2')-f({\bf V}_1)f({\bf V}_2)\right],
\label{A1}
\eeq
where
\beq
{\bf V}_1'={\bf V}_1-\left({\bf
g}\cdot\widehat{\bbox{\sigma}}\right)\widehat{\bbox{\sigma}}, 
\quad
{\bf V}_2'={\bf V}_2+\left({\bf
g}\cdot\widehat{\bbox{\sigma}}\right)\widehat{\bbox{\sigma}}. 
\label{A2}
\eeq
Here, ${\bf g}\equiv {\bf V}_1-{\bf V}_2$ is the relative velocity and
$\widehat{\bbox{\sigma}}$ is a unit vector lying on the scattering plane and
 making  an angle equal to
$\frac{1}{2}(\pi-\theta)$ with the vector ${\bf g}$.

Given an arbitrary function $H({\bf V}_1)$, standard manipulations yield
\beq
\int d{\bf V}_1\, H({\bf V}_1) Q[{\bf V}_1|f,f]=
\frac{1}{2}\int d{\bf V}_1\int d{\bf V}_2\int d\Omega\, K(\theta)f({\bf
V}_1)f({\bf V}_2)\left[H({\bf V}_1')+H({\bf
V}_2')-H({\bf V}_1)-H({\bf V}_2)\right].
\label{A3}
\eeq
In the particular case of $H({\bf V}_1)=m{\bf V}_1{\bf V}_1$,
\beq
H({\bf V}_1')+H({\bf
V}_2')-H({\bf V}_1)-H({\bf V}_2)=m\left({\bf
g}\cdot\widehat{\bbox{\sigma}}\right)\left[2\left({\bf
g}\cdot\widehat{\bbox{\sigma}}\right)\widehat{\bbox{\sigma}}\widehat{\bbox{\sigma}}
-\widehat{\bbox{\sigma}}
{\bf g}-{\bf g}\widehat{\bbox{\sigma}}\right].
\label{A4}
\eeq
Taking into account the identities
\beq
\int d\Omega\, K(\theta)\left({\bf
g}\cdot\widehat{\bbox{\sigma}}\right)\widehat{\bbox{\sigma}} =B_1 {\bf g},
\label{A5}
\eeq
\beq
\int d\Omega\, K(\theta)\left({\bf
g}\cdot\widehat{\bbox{\sigma}}\right)^2\widehat{\bbox{\sigma}}
\widehat{\bbox{\sigma}}=B_2 {\bf g}{\bf g}-\frac{B_1-B_2}{d-1}\left({\bf g}{\bf
g}-g^2{\sf I}\right),
\label{A6}
\eeq
where ${\sf I}$ is the $d\times d$ unit tensor and
\beq
B_k\equiv \int d\Omega\, K(\theta) \sin^{2k}\frac{\theta}{2},
\label{A7}
\eeq
we finally have
\beqa
\int d{\bf V}_1\, m {\bf V}_1{\bf V}_1 Q[{\bf V}_1|f,f]&=&-\frac{d}{d-1}(B_1-B_2)
\int d{\bf V}_1\int d{\bf V}_2f({\bf V}_1)f({\bf V}_2)m\left({\bf g}{\bf
g}-\frac{1}{d}g^2{\sf I}\right)\nonumber\\
&=&-\frac{2d}{d-1}(B_1-B_2)n\left({\sf P}-p{\sf I}\right).
\label{A8}
\eeqa
This allows us to identify the effective collision frequency of Eq.\ 
(\ref{b12}) as $\nu=n(B_1-B_2)2d/(d-1)$.
\section{Derivation of Eq.\ \lowercase{(\protect\ref{c2})}}
\label{appB}
Let us start rewriting Eq.\ (\ref{A3}) as
\beq
I[H]\equiv\int d{\bf V}_1\, H({\bf V}_1) Q[{\bf V}_1|f,f]=
\int d{\bf V}_1\int d{\bf V}_2\int d\Omega\, K(\theta)f({\bf
V}_1)f({\bf V}_2)\left[H({\bf V}_1')-H({\bf V}_1)\right],
\label{B1}
\eeq
where $H({\bf V}_1)$ is an arbitrary function. Now, according to the scattering
law (\ref{c1}), $\left({\bf
g}\cdot\widehat{\bbox{\sigma}}\right)=g\sin(\epsilon/2)$, so that ${\bf
V}_1'-{\bf V}_1 \sim \epsilon$. This justifies the approximation
\beq
H({\bf V}_1')-H({\bf V}_1)\simeq -\frac{\partial H({\bf V}_1)}{\partial V_{1i}}\left({\bf
g}\cdot\widehat{\bbox{\sigma}}\right)\widehat{\sigma}_i+\frac{1}{2}
\frac{\partial^2 H({\bf V}_1)}{\partial V_{1i}\partial V_{1j}}\left({\bf
g}\cdot\widehat{\bbox{\sigma}}\right)^2\widehat{\sigma}_i\widehat{\sigma}_j.
\label{B2}
\eeq
Consequently,
\beq
I[H]\to\int d{\bf V}_1\int d{\bf V}_2\,f({\bf V}_1)f({\bf V}_2)\left\{-B_1
\frac{\partial H({\bf V}_1)}{\partial V_{1i}} g_i+ \frac{1}{2}
\frac{\partial^2 H({\bf V}_1)}{\partial V_{1i}\partial V_{1j}}\left[B_2
{g}_i{g}_j-\frac{B_1-B_2}{d-1}\left({g_i}
g_j-g^2\delta_{ij}\right)\right]
\right\},
\label{B3}
\eeq
where we have made use of (\ref{A5}) and (\ref{A6}). On the other hand, $\lim_{\epsilon\to
0}B_2=0$, while $B_1\to (\nu/n)(d-1)/2d$ remains constant. Integrating over
${\bf V}_2$ in Eq.\ (\ref{B3}), we have
\beq
I[H]\to\frac{d-1}{2d}\nu\int d{\bf V}_1\, H({\bf V}_1)\frac{\partial}{\partial V_{1i}}
\left[V_{1i}-\frac{1}{2(d-1)}\frac{\partial}{\partial
V_{1j}}\left(V_{1i}V_{1j}-V_1^2\delta_{ij}+\frac{P_{ij}-dp\delta_{ij}}{mn}\right)\right]f({\bf
V}_1),
\label{B4}
\eeq
where we have integrated by parts. Since $H({\bf V}_1)$ is arbitrary, it follows
that
\beq
Q[{\bf V}_1|f,f]\to \frac{d-1}{4d}\nu 
\frac{\partial}{\partial V_{1i}}
\left[V_{1i}+\frac{1}{d-1}\left(\frac{dp\delta_{ij}-
P_{ij}}{mn}+V_1^2\delta_{ij}-V_{1i}V_{1j}\right)
\frac{\partial}{\partial
V_{1j}}\right]f({\bf
V}_1),
\label{B5}
\eeq


\end{document}